\documentclass[pra,twocolumn,showpacs,superscriptaddress,floatfix]{revtex4}
\usepackage{amsmath,amssymb,amsfonts,graphicx,bbm}
\newcommand{\ket}[1]{\vert #1 \rangle} \newcommand{\bra}[1]{\langle #1 \vert}

\newcommand{\ketbra}[2]{\left\vert #1 \rangle \! \langle #2 \right\vert}

\newcommand{\average}[1]{\left \langle #1  \right\rangle}
\newcommand{\be}{\begin{equation}}
\newcommand{\ee}{\end{equation}}
\newcommand{\smm}{{\scriptscriptstyle M}}
\newcommand{\bae}{\begin{eqnarray}} \newcommand{\eae}{\end{eqnarray}}

\newcommand{\bmlambda}{\boldsymbol \lambda}
\newcommand{\HH}{{\hbox{\small HH}}}
\newcommand{\HV}{{\hbox{\small HV}}}
\newcommand{\VH}{{\hbox{\small VH}}}
\newcommand{\VV}{{\hbox{\small VV}}}

\def\({\left(} \def\){\right)} \def\[{\left[} \def\]{\right]}
  
\def\Tr{\hbox{Tr}}

\begin{document}
\title{Optimal estimation of entanglement in optical qubit systems}
\author{Giorgio Brida}
\affiliation{INRIM, I-10135, Torino, Italy}
\author{Ivo P. Degiovanni}
\affiliation{INRIM, I-10135, Torino, Italy}
\author{Angela Florio}
\affiliation{INRIM, I-10135, Torino, Italy}
\author{Marco Genovese}\email{m.genovese@inrim.it}
\affiliation{INRIM, I-10135, Torino, Italy}
\author{Paolo Giorda}\email{giorda@isi.it}
\affiliation{ISI Foundation, I-10133, Torino, Italy}
\author{Alice Meda}
\affiliation{INRIM, I-10135, Torino, Italy}
\author{Matteo G. A. Paris}\email{matteo.paris@fisica.unimi.it}
\affiliation{Dipartimento di Fisica, Universit\`a degli Studi di Milano,
I-20133 Milano, Italy}
\author{Alexander P. Shurupov}
\affiliation{INRIM, I-10135, Torino, Italy}
\date{\today}
\begin{abstract}
We address the experimental determination of entanglement for
systems made of a pair of polarization qubits. We exploit quantum
estimation theory to derive optimal estimators, which are then
implemented to achieve ultimate bound to precision. In particular,
we present a set of experiments aimed at measuring the amount of
entanglement for states belonging to different families of pure and
mixed two-qubit two-photon states. Our scheme is based on visibility
measurements of quantum correlations and achieves the ultimate
precision allowed by quantum mechanics in the limit of Poissonian
distribution of coincidence counts. Although optimal estimation of
entanglement does not require the full tomography of the states we
have also performed state reconstruction using two different sets of
tomographic projectors and explicitly shown that they provide a less
precise determination of entanglement. The use of optimal estimators
also allows us to compare and statistically assess the different noise
models used to describe decoherence effects occuring in the generation of
entanglement.
\end{abstract}
\maketitle
\section{Introduction}
The sort of quantum correlations captured by the notion of
entanglement represents a central resource for quantum information
processing. Therefore, the precise characterization of entangled states is a
crucial issue for the development of quantum technologies. In fact,
quantification and detection of entanglement have been extensively
investigated, see \cite{ren1,ren2,ren3} for a review, and different
approaches have been developed to extract the amount of entanglement
of a state from a given set of measurement results
\cite{bayE,Wun09,Eis07,KA06}. Of course, in order to evaluate the
entanglement of a quantum state one may resort to full quantum state
tomography \cite{LNP} that, however, becomes impractical in higher
dimensions and may be affected by large uncertainty \cite{TE94,AS2011}.
Other methods, requiring a reduced number of observables, are based
on visibility measurements \cite{Jae93}, Bell' tests \cite{c2,w3},
entanglement witnesses \cite{h1,t2,G3,w,w1} or are related to
Schmidt number \cite{pas,fed,f}. Many of them has been implemented
experimentally \cite{mg,wei,ser,nos,ch,mat}, also in the presence of
decoherence effects \cite{buc06,dav07}.
\par
As a matter of fact, any quantitative measure of entanglement corresponds to a
nonlinear function of the density operator and thus it cannot be associated to
a quantum observable. As a consequence, ultimate bounds to the precision of
entanglement measurements cannot be inferred from uncertainty relations. Any
procedure aimed to evaluate the amount of entanglement of a quantum state is
ultimately a parameter estimation problem, where the value of entanglement is
indirectly inferred from the measurement of one or more proper observables
\cite{EE08}. An optimization problem thus naturally arises when one looks for
the ultimate bounds to precision, {\em i.e.} the smallest value of the entanglement
that can be discriminated according to quantum mechanics, and tries to determine
the optimal measurements achieving those bounds. This optimization problem may
be properly addressed in the framework of quantum estimation theory
\cite{qet1,qet2,qet3}, which provides analytical tools to find the optimal
measurement and to derive ultimate bounds to the precision of entanglement
estimation. In particular, being entanglement an intrinsic property of quantum
states, we adopt local quantum estimation theory and look for optimal estimators maximizing the Fisher information \cite{EE08,LQE}.
\par
In this paper, we address experimental determination of entanglement for
two-qubit  optical systems and apply quantum estimation theory to derive
optimal estimators and ultimate bound to precision. This technique has
been successfully applied in \cite{EEE}  to estimate the entanglement of
a pair of polarization qubit with the ultimate precision allowed by
quantum mechanics. Here we refine and extend the results of \cite{EEE}
in two directions: On the one hand we present a set of experiments aimed
at estimating the amount of entanglement of a larger class of families
of two-qubit mixed photon states.  On the other hand, we have performed
full state reconstruction using two different tomographic sets of
projectors in order to show explicitly that the evaluation of
entanglement from the knowledge of the reconstructed density matrix
provides a less precise determination.  In our scheme entanglement, is
evaluated through visibility measurements and estimators are built by a
suitable combination of coincidence counts with different settings.
Those estimators turn out to be optimal and to provide estimation with
the ultimate precision in the limit of Poissonian distribution of
coincidence counts. In addition, we demonstrate experimentally that optimality
is robust against deviation from the Poissonian behaviour.
Our approach allows entanglement estimation at
the quantum limit, and it is also useful to compare
different noise models using only information extracted from
experimental data.
\par
The paper is structured as follows. In the next Section we briefly
review the basic notions of local estimation theory, whereas in
Section \ref{sec:models.theory} we apply them to
estimation of entanglement of states belonging to two relevant
families of mixed states. Section \ref{sec:exp.app} describes in details the
experimental apparatus used to demonstrate our theoretical results, which are
described in the Section \ref{sec:estimators}.  A detailed discussion of
the experimental results is given in Section \ref{sec:results}, whereas
Section \ref{sec:outro} closes the paper with some concluding remarks.
\section{Local quantum estimation theory}
We now give the basis ingredients for the local estimation theory
 starting with the classical case. Suppose
we have a set of parameters ${\bmlambda}=(\lambda_1,\cdots,\lambda_n)\in
\Lambda \subseteq \mathbb{R}^n$
labelling different states of the physical system of interest. A
statistical model of our system is a set of probability distributions
$S=(p_{{\bmlambda}}(x)|{\bmlambda}\in \Lambda)$ such that $\Omega$ is
the sample space of the random variable $x$.  The fundamental question in
estimation theory is how to optimally estimate the unknown true values
of the parameters ${\bmlambda}$ given a sequence of outcomes of
measurement on the system $\{x_1,\cdots,x_\smm\}$. From an geometrical
information perspective, this problem was first treated by Fisher
who introduced for the case $N=1$  the now called Fisher
information metric $F({\bmlambda})$:
 \bae\label{Eq.: FisherInfo}
 [F({\bmlambda})]_{ij}&=&\int_{\Omega}\!\! dx\, p_{{\bmlambda}}(x)\,
 \partial_i \log p_{{\bmlambda}}(x)\,
 \partial_j \log p_{{\bmlambda}}(x) =\nonumber \\
 &=&\int_\Omega\!\! dx\, \frac{\partial_i p_{{\bmlambda}}(x) \partial_j
 p_{{\bmlambda}}(x) }{p_{{\bmlambda}}(x)}
 \eae
where $\partial_i\equiv\partial_{\lambda_i}$. 
$F({\bmlambda})$ is a positive definite matrix that represents a
metric on the parameter space $\Lambda$ and whose information
geometric content is given by the best resolution with which one can
distinguish neighbouring points in the parameter space. The Fisher
information metric is additive, therefore for a sequence of
independent and identically distributed measurements with outcomes
$\{x_1,\cdots,x_\smm\}$, $F^\smm({\bmlambda})=M F({\bmlambda})$. The
next step in the estimation theory requires the introduction of the
concept of estimator; the latter is any  algorithm or rule of inference,
which allows one to extract a value for the unknown parameters on the
basis of the sole knowledge acquired via the measurement process, i.e. the
sequence of outcomes $\{x_1,\cdots,x_\smm\}$. We say that the random variable
$\hat{\bmlambda}: \Omega^\smm \rightarrow \Lambda$ is an unbiased
estimator if $E[{\hat \bmlambda}]={\bmlambda}$ i.e., its expected value
coincides with the true value of the parameter(s). The ultimate bound on
the precision with which one can estimate the parameters ${\bmlambda}$ is
given by the Cramer-Rao theorem , which can be stated in terms of the
covariance matrix $\hbox{Cov}[\hat{ \bmlambda}]_{ij}=E[\hat{\lambda}_i
\hat{\lambda}_j]-E[\hat{\lambda}_i]E[\hat{\lambda}_j]$ as:
 \be\label{Eq.: ClassCramerRao}
  \hbox{Cov}[\hat{\bmlambda}]\ge \frac{1}{M} F({\bmlambda})^{-1}.
 \ee
In particular, for a single parameter the inequality reads
$$\hbox{Var}[\hat{\lambda}]\ge\frac1{M F(\lambda)}\,,$$ i.e. the variance
of the estimator, and therefore the precision of any estimation
procedure, cannot be smaller than the inverse of the Fisher information
times the number of repeated measurements. In the general case, the
inequality for the variance of each of the parameters, i.e.
$$\hbox{Var}[\hat{\lambda}_i]\ge \frac1M [F({\bmlambda})^{-1}]_{ii}\,,$$ holds
only at fixed values of the others parameters.
\par
The previous results can be extended to the quantum realm, also taking
into account all the possible measurements that one can implement on the
systems. The quantum statistical model is given by a set of density operators
depending on the parameters ${\bmlambda}$: $S=\{\rho_{\bmlambda}|{\bmlambda}\in
\Lambda\}$. A measurement corresponds to a Positive Operator Valued Measure
(POVM), i.e. a set of positive operators $\mathcal{E}=\{E_i\}$ such that
$\sum_i E_iE_i^\dagger=\openone$ and such 
that $p_{{\bmlambda}}(i)=\Tr[E_i\rho_{\bmlambda}]$ is the probability of having the
$i$-th outcome. The Fisher information matrix $F_{\mathcal{E}}(\bmlambda)$ in Eq.
(\ref{Eq.: FisherInfo}) for a specific measurement process $\mathcal{E}$ can
then be written in terms of the classical probabilities $p_{{\bmlambda}}(i)$.
What is now specific to the quantum estimation process is that the optimization
over measurement processes $\mathcal{E}$ may be carried out. The problem has
been solved in terms of the inequality ($A > B$ means that $A-B$ is a
positive matrix)
 \be
 F_{\mathcal{E}}(\bmlambda)\le H(\bmlambda)
 \ee
that states that the Fisher information of any measurement process is upper
bounded by the Quantum Fisher information $H(\bmlambda)$ (QFI). The latter
is an $n\times n$ positive definite real matrix which can be expressed in terms
of a set of $n$ positive, zero mean operators called symmetric logarithmic
derivatives (SLD) $L_i$, each satisfying the following partial differential
equation
 \be\label{Eq.: SDLi}
 \partial_i \rho_{{\bmlambda}}=\frac{1}{2}(L_i\rho_{\bmlambda}+\rho_{\bmlambda}L_i)
 \ee
In particular, if one expresses
the density matrix in its spectral decomposition
 \be
 \rho_{\bmlambda}=\sum_i p_i \ketbra{\psi_i}{\psi_i},
 \ee
the SLD pertaining to the $i$-th parameter is
 \be
 L_i  = 2 \sum_{n,m}
 \frac{\langle\psi_n | \partial_i \rho_{\bmlambda}|\psi_m\rangle }
 {p_n + p_m} |\psi_n\rangle\langle\psi_m|,
 \ee
where
 \begin{align}
 \partial_i \rho_{\bmlambda}=&\sum_n \partial_i p_n \ketbra{\psi_n}{\psi_n}+ \\
 & +
 \sum_n p_n (\ketbra{\partial_i \psi_n}{\psi_n}+\ketbra{\psi_n}{\partial_i\psi_n})
\,,\notag
\end{align}
accounts for the dependence of both the eigenvalues and the eigenvectors
on the set of parameters ${\bmlambda}$. In terms of the $L_i$'s the elements
of the QFI can be written as:
 \be
 [H(\bmlambda)]_{ij}=\Tr\left[\rho_{\bmlambda}\,\frac{L_iL_j+L_jL_i}{2}\right].
 \ee
By using the spectral decomposition of $\rho_{\bmlambda}$, the QFI can be
expressed in terms of the partial derivatives of the eigenvalues and of the
eigenvectors as:
 \begin{align}\label{Eq.: QFIeig}
 &[H(\bmlambda)]_{ij}=
 \sum_n \frac{(\partial_i p_n) (\partial_j p_n)}{p_n} +
 \sum_{n,m}
 \frac{(p_n-p_m)^2}{p_n + p_m} \times \\
 & \times \Big(\langle \psi_n|\partial_i \psi_m\rangle
 \langle \partial_j \psi_m| \psi_n \rangle +
\langle \psi_n|\partial_j \psi_m\rangle
 \langle \partial_i \psi_m| \psi_n \rangle \Big)\,.\nonumber
 \end{align}
\section{Estimation of entanglement for two-qubit systems}
\label{sec:models.theory}
We now apply the formalism described in the previous Section to obtain
explicitly the ultimate bound to precision on the estimation of
entanglement for two relevant statistical models, i.e. for two
families of two-qubit states that will be used in the following.
\subsection{The decoherence model}
The first statistical model we are going to deal with
corresponds to the set of the states described by the following
two-parameter family of density operators
\be\label{Eq.: DM.deco}
\varrho = p\ketbra{\psi}{\psi}+(1-p)D,
\ee%
where \be\label{Eq.: state_with_q}
 \ket{\psi}=\sqrt{q}\,\ket{\HH}+\sqrt{1-q}\,\ket{\VV}
\ee
represents a pure polarization two-photon state with horizontal $H$
and vertical $V$ polarization, and $D = q\,\ketbra{\HH}{\HH} +
(1-q)\,\ketbra{\VV}{\VV}$ describes a mixed contribution coming from
the decoherence of $|\psi\rangle$, $p\in [0,1]$. We will refer to
this set as the decoherence model for $|\psi\rangle$.  For the state
$\varrho$, both the two non zero eigenvalues
$$\lambda_{\pm}=(1\pm\sqrt{1-4(1-p^2)q+4(1-p^2)q^2})\,$$
and their respective eigenvectors
\begin{align}
{\bf{v}}_{\pm}&= \frac{1}{\sqrt{N_\pm}}\left\{ -f_{\pm}(p,q), 0, 0, g(p,q)\right\}\\
N_\pm &=\sqrt{g^2(p,q)\pm f^2_{\pm}(p,q)} \notag\\
f_{\pm}(p,q)&=1-2q\pm\sqrt{1-4(1-p^2)q+4(1-p^2)q^2} \notag \\
g(p,q)&=2 p \sqrt{q(1-q)} \,\notag
\end{align}
depend on the
parameters $p,q$. The straightforward calculations of the partial derivatives
in Eq. (\ref{Eq.: QFIeig}) show that both the eigenvalues and the eigenvectors
contribute to the diagonal and off-diagonal terms of the QFI.
However, the sum of the different contributions results in a simplified
expression, and the QFI
\be
H(p,q)=\mbox{diag}\left(\frac{4(1-q)q}{1-p^2},\frac{1}{q-q^2}\right)
\ee%
is diagonal. From this expression we see that the variance on any
estimator $\hat{q}$ for the parameter $q$ is independent on the mixing
parameter $p$ and is bounded, apart from the statistical scaling, by
the inverse of corresponding element
of the QFI matrix
$$\hbox{Var}[\hat{q}]\ge \frac{q(1-q)}{M}\,.$$ The lower bound is maximal
in correspondence of $q=1/2$, i.e. when the state $\ket{\psi}$ is maximally
entangled. We are now interested in estimating the value of entanglement of
the overall state $\varrho$. To this aim we remind that the negativity of
entanglement
defined as
 \be\label{Eq.: NegativityDefinition}
 \epsilon = \vert\vert \varrho^{T_A} \vert\vert_1 - 1
 \ee
is a good measure of entanglement for two
qubit systems. In Eq. (\ref{Eq.: NegativityDefinition})
$T_A$ denotes partial transposition with respect to system $A$, and
$\vert\vert...\vert\vert_1$ is the trace norm. Entanglement
negativity for states belonging to the decoherence model
is given by
\be\label{Eq.: NegativityValue}
\epsilon=2 p\sqrt{(1-q)q}\,.
\ee
In order to reexpress the QFI in terms of the negativity we make the change of
variable $p\rightarrow p, q\rightarrow (p-\sqrt{p^2-\epsilon})/2p$; the QFI
changes according to the Jacobian of the transformation and the lower bound to
the covariance matrix of the estimators $\hat{p},\hat{\epsilon}$ now reads:
 \begin{align}
 \hbox{Cov}[\hat{p},\hat{\epsilon}]&\ge H^{-1}(p,\epsilon) \\
&=
  \left(
   \begin{array}{cc}
  p^2(1-p^2)\,\epsilon^{-2} & p(1-p^2)\,\epsilon^{-1} \\
    p(1-p^2)\,\epsilon^{-1}& 1-\epsilon^2 \\
   \end{array}
  \right)\notag
 \end{align}
From this expression we see that the lower bound for the variance of any
estimator $\hat{\epsilon}$  of the negativity of the state
$\varrho$ is independent on $p$ and is minimal in case of maximal
entanglement
 \be\label{Eq.: He}
  \hbox{Var}[\hat{\epsilon}]\ge \frac1M(1-\epsilon^2)\,.
 \ee
\subsection{The Werner model}
A second statistical model of interest for our analysis corresponds to
the set of states described by the following two-parameter family of density
operator
\be\label{Eq.: DM.werner}
\varrho^\prime =p \ketbra{\psi}{\psi}+ \frac{1-p}{4}\openone \otimes \openone
\,.
\ee
The states of Eq. (\ref{Eq.: DM.werner}) are obtained by depolarizing
the pure entangled state $\ket{\psi}$. We will refer to this family as
the Werner model for $|\psi\rangle$. As in the previous example upon
varying the parameter $p$ we may tune the purity of the state, whereas
the amount of entanglement depends on both parameters. The eigenvalues
of $\varrho^\prime$ depends only on $p$, whereas the eigenvectors depends
only on $q$. The QFI matrix is thus given by the diagonal form
\be
H(p,q)=\mbox{diag}\left \{ \frac{3}{1+(2-3p)p},\, \,
\frac{p^2}{q(1-q)(1+p)} \right \}\,
\ee
and the inverses of the diagonal elements correspond
to the ultimate bounds to ${\mathrm{Var}}(\hat{p})$ and ${\mathrm{Var}}(\hat{q})$
for any estimator of $p$ and $q$, either at fixed value of the other
parameter or in a joint estimation procedure. Entanglement of Werner
states may be evaluated in terms of negativity,
\be
\epsilon=\max \left\{0,
\frac12 \left[p\left(1+4\sqrt{q(q-1)}\right)-1\right]\right\}
\label{wen}
\:,
\ee
which implies that Werner states are entangled for
$$[1+ 4\sqrt{q (1 - q)}]^{-1} < p < 1.$$
Upon inverting Eq. (\ref{wen}) for $p$ or $q$ we may parametrize the Werner
states using $(p,\epsilon)$ and evaluate the QFI matrix $H(p,\epsilon)$, their
inverses and, in turn, the corresponding bounds to the precision of
entanglement estimation. The main result is that the ultimate bound to the
variance, depend only very slightly on the other free parameter ($q$ or $p$).
In other words, estimation procedures performed at fixed value of $p$ or $q$
respectively show different precision, but the differences are negligible in
the whole range of variations of the parameters.  We do not report here the
analytic expression of the inverse QFI at fixed $p$ or $q$, which is quite
cumbersome. However, as it can be easily checked, we note that the bound on the
variance on $\hat{\epsilon}$ that can be derived by the expression of
$H(p,\epsilon)^{-1}$ simply coincides to first order with the bound in Eq.
(\ref{Eq.: He}) already evaluated for the decoherence model. We therefore use in
the following, also for the Werner model, the bound given in Eq. (\ref{Eq.: He}).
It can be shown that, for the set of values of $p$ that will be
relevant for our experimental analysis, this approximation
is negligible with respect to all the other sources of
uncertainty.
\section{Experimental apparatus}\label{sec:exp.app}
\begin{figure}[h]
\includegraphics[width=0.95\columnwidth]{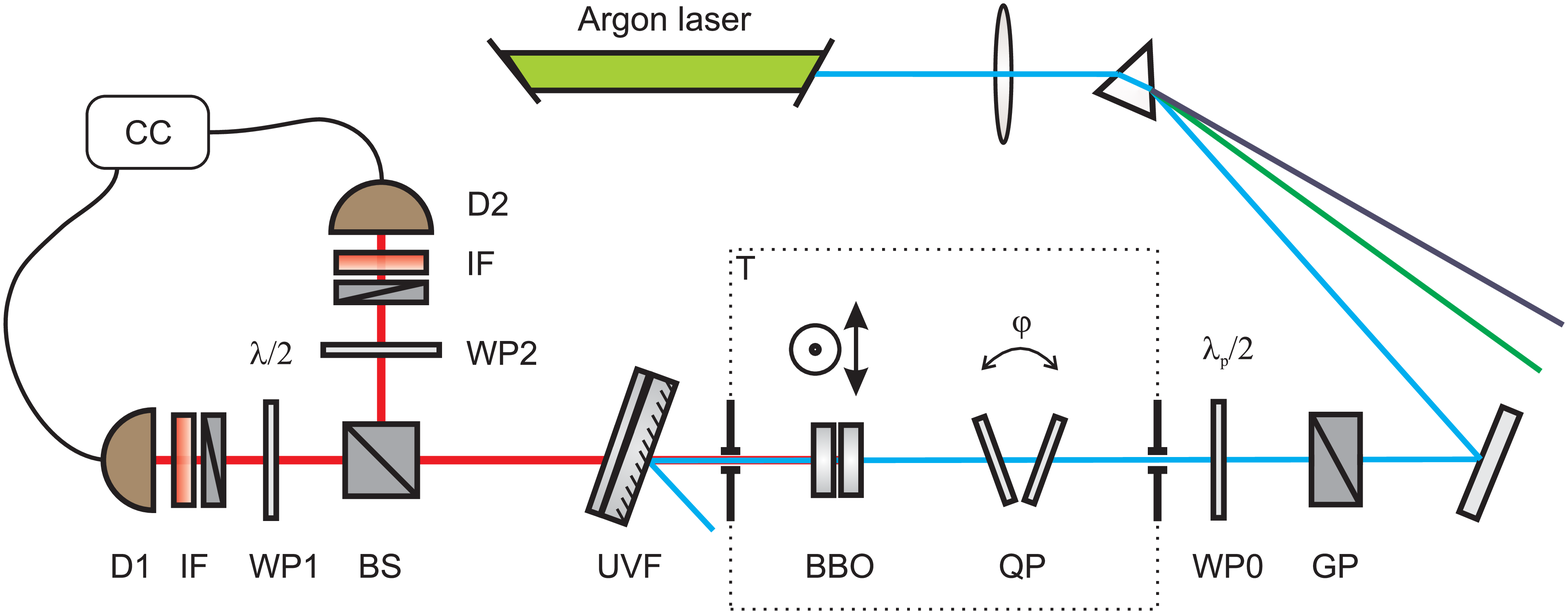}
\caption{(Color online) Experimental setup to generate polarization entangled
two-photon states with variable entanglement and to estimate its value with the
ultimate precision allowed by quantum mechanics. A continuous wave Argon pump
laser beam with wavelength $\lambda$ = 351.1 nm is filtered with a dispersion
prism and then passes through a Glan-Thompson prism and a half-wave plate WP0
that rotates the polarization by an angle $\phi$. PDC light is generated by two
thin type-I BBO crystals ($l$ = 1 mm). After the crystals the pump is stopped by
a filter (UVF), and the biphoton field is split on a nonpolarizing 50-50 beam
splitter (BS). Then it passes through half-wave plates (WP1, WP2) and
interference filters (IF), centered at the degeneracy 702 nm. Finally the
biphotons are focused on commercial single photon detectors (D1, D2).}
\label{f:setup}
\end{figure}

The family of entangled states, investigated in our work, is
constituted by polarization entangled states of the field obtained
by coherently superimposing two orthogonally polarized type-I
parametric downconversion emissions (PDC), as schematically depicted
in Fig.~\ref{f:setup}. The linear horizontal polarization of an
argon laser beam, at wavelength $\lambda$ = 351 nm filtered by
dispersion prism and Glan-Thompson prism (GP), is rotated at angle
$\phi$ by using half-waveplate (WP0). It is fundamental for our
application that only the laser line $\lambda$ = 351.1 nm is used. For
this reason we have introduced in the setup a prism as wavelength
selector for eliminating wavelengths other than $\lambda$ = 351.1 nm.
In particular the closest one at $\lambda$ = 351.4 nm, which could
realize an unwanted phase-matching condition in our PDC
setup. Then, the laser beam is addressed to a pair of non-linear
beta barium borate (BBO) crystals ($l$ = 1 mm), having optical axis in
orthogonal planes, where PDC process occurs, resulting in creation
of biphotons with orthogonal polarization \cite{kw,nos1}. Upon
changing the polarization of the UV pump, we change the amount of
PDC light, generated by each crystal. For example, PDC occurs only
in crystal one if the polarization of the pump beam is horizontal,
while for having a balanced PDC process in both crystals we have set
the angle $\phi$ at $45^\circ$, having diagonal polarization of pump beam.
\par
In order to compensate phase shifts, due to ordinary and extraordinary path
in the crystals, we tilt the quartz plates QP, introduced between the halfwave
plate WP0 and BBO crystals, at angle $\varphi$, thus fixing the relative phase
between biphoton components generated in first and second crystal.
\par
In order to maintain stable the phase-matching conditions, BBO
crystals and QP are placed in a closed aluminium box
internally covered by polystyrene used as thermic insulator.
The box is equipped with a controlled heating system with a standard
feedback circuit. We have experimentally verified that the
temperature stabilization system ensures appropriate control on the
phase shift. After the box the pump is stopped by an ultraviolet
filter (UVF), and the biphoton field is split on a non-polarizing
50-50 beam splitter (BS). With the postelection performed by a
coincidence count circuit (CC), we can refer to our state as an
optical ququart \cite{kulik}, which is entangled in two variables:
polarization and spatial mode.
\par
In ideal conditions the output state is described by the pure
state
 \be\label{Eq.: state_with_phase}
 \ket{\psi_{\phi\varphi}} = \cos\phi\ket{\HH} + \sin\phi e^{i\Phi(\varphi)}
 \ket{\VV}
 \ee
where $\phi/2$ is rotation angle of pump halfwaveplate WP0 and
$\Phi(\varphi)$ corresponds to phase shift between pair of
horizontal photons created in the first crystal and pair of vertical
photons from the second crystal. After passing the half-waveplates
(WP1,WP2) in each spatial mode, the biphoton field is projected into
a linear vertical polarization state by means of Glan-Thompson
polarizers. Phase plates WP1 and WP2 are mounted on precision
rotation stages with high resolution and fully motor controlled,
that allow rotating the polarization of the beams in the course of
measurement process. Spectral selection is performed by interference
filters (IF) with central wavelength $\lambda$ = 702 nm and FWHM = 3 nm.
Short focal lenses collimate resulting biphoton field into single
photon avalanche detectors (D1, D2). Electrical signal from
detectors is used by coincidence count scheme (CC) with time window
$\tau$ = 1 ns.
\par
The measurements performed at the output are described as projection of
state into factorized linearly-polarized two-photon state:
 \be\label{Eq.: projectors operators}
 \Pi_x(\alpha,\beta)=
 \ket{\alpha+s\frac{\pi}2}\bra{\alpha+s\frac{\pi}2}\otimes
 \ket{\beta+s'\frac{\pi}2}\bra{\beta+s'\frac{\pi}2}
 \ee
where $x=\{s+2s'\}$, $s,s'=0,1$.
\begin{figure}[h]
\includegraphics[width=0.95\columnwidth]{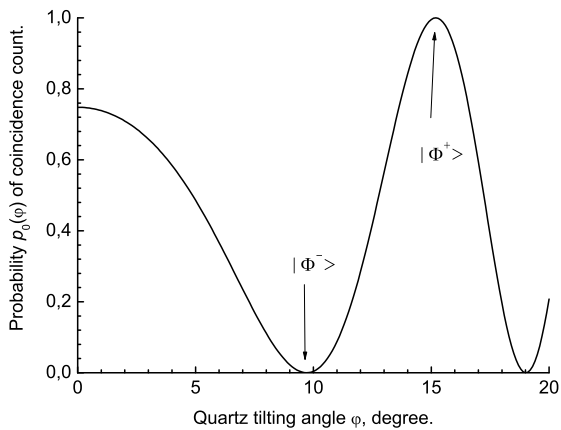}
\caption{Probability of coincidence counts while performing projection
measurement $\Pi_0(\frac{\pi}4,\frac{\pi}4)$ on state
$\ket{\psi_{\phi\varphi}}$ having $\phi=\frac{\pi}4$ as function of quartz
plates tilting angle $\varphi$.} \label{f:delta_phase}
\end{figure}
\par
In Fig.~\ref{f:delta_phase} we show the dependence of the probability of the
coincidence counts $$p_0(\varphi) = \bra{\psi_{\tfrac{\pi}{4}\varphi}}
\Pi_0(\frac\pi4,\frac\pi4) \ket{\psi_{\tfrac{\pi}{4}\varphi}}\,,$$ as function of quartz
plates QP tilting angle $\varphi$. The maximum of this curve corresponds to
phase shift between photon pairs $\Phi(\varphi_\smm)=0$ and the output state is the
Bell maximally entangled state $$\ket{\Phi^+}\equiv\ket{\psi_{\frac{\pi}{4}\varphi_\smm}}
\propto \ket{\HH} + \ket{\VV}\,,$$ while the minimum of that curve
corresponds to the maximally entangled state
$$\ket{\Phi^-}\equiv\ket{\psi_{\frac{\pi}{4}\varphi_m}}
\propto \ket{\HH} - \ket{\VV}\,.$$ In this work we
have fixed the tilting angle of quartz plates to have zero phase shift, thus, the
family of states in Eq. (\ref{Eq.: state_with_phase}) reduce to the one of
Eq. (\ref{Eq.: state_with_q}) where $q=\cos^2(\phi)$.
\section{Entanglement estimators} \label{sec:estimators}
In order to estimate the entanglement content of the states
produced by the experimental set up described in the previous Section,
one has to choose an estimator $\hat{\epsilon}$ to extract the
value of entanglement from the experimental data. We will compare
three different approaches: two are based on full tomography of
the polarization two-photon and one is based on implementing
the optimal estimator able to saturate the ultimate bound derived via
the QFI.
\par
Quantum state tomography is an experimental procedure providing full density
matrix reconstruction of a quantum system. This is realized by means of a set
of measurements performed on an ensemble of identical quantum systems
\cite{LNP}. For a quantum state belonging four-dimensional Hilbert space at
least 16 linearly independent measurements are needed to reconstruct full
density matrix and, typically, each measurement corresponds to a local
projection of the input two-qubit state. To be able to perform this set of 16
linearly independent measurements we added a quarter-waveplate in each
measurement arm just before the half-waveplates (WP1,WP2). The first used
tomographic protocol (J16) \cite{KB00,Ja01} involves projective measurements
performed directly on some components of the Stokes vector. In particular, the
measurement set corresponds to projection onto polarizations
HH, HV, VV, VH, RH, RV, DV, DH, DR, DD,
RD, HD, VD, VL, HL, RL, where H, V, R, L, D, denotes horizontal, vertical,
right and left circular and $45^\circ$ diagonal polarizations, respectively.
Here, for example, the measurement setting HR means measuring horizontal
polarization on the first qubit and right circular polarization on the second
qubit. Another approach \cite{bog04,reh04} involves local projection
of each qubit symmetrically placed on Poincare sphere. Extension of this
method to four-dimensional case (R16) allows obtaining higher fidelity of the
reconstructed states~\cite{Bur08,OurTomo2010} with respect to the previous one.
Once the density matrix of the generated state has been reconstructed, the
negativity of the state can be evaluated inserting the reconstructed matrix
elements in Eq.
(\ref{Eq.: NegativityDefinition}). The precision the tomographic estimation
of entanglement is limited by the uncertainties on the matrix elements. The
overall uncertainty on the estimated value of entanglement may be evaluated
by error propagating. In the following, after describing the implementation
of optimal measurement, we will compare its precision with that of
tomographic estimation.
\par
We first start to briefly describe the estimator for the class of states
defined by Eq. (\ref{Eq.: DM.deco}). As already described in \cite{EEE},
an optimal estimator of the entanglement can be found by noticing that
the expressions of the probabilities $p_x(\epsilon; \alpha,\beta)=
\hbox{Tr}[\varrho\: \Pi_x(\alpha,\beta)]$
obtained by the projection of the state $\varrho$ on measurement operators
in Eq. (\ref{Eq.: projectors operators}) with
$x=0,1,2,3$, allows  writing the following set of unbiased estimators%
\be\label{eq:estimator}
 \hat\epsilon(\alpha,\beta) = \frac{V(\alpha,\beta) -
 \cos(2\alpha)\cos(2\beta)}{\sin(2\alpha)\sin(2\beta)},
\ee%
where $V(\alpha,\beta)=p_0-p_1-p_2+p_3$ is the expected value of
two-qubit quantum correlations (QC).
Furthermore, the estimators corresponding to the measurement angles
$\alpha,\beta=\pm\pi/4$ are optimal, as can be seen by evaluating
the Fisher information $$F_\epsilon(\alpha,\beta) = \sum_x
p_x(\epsilon;\alpha,\beta)[\partial_\epsilon\log
p_x(\epsilon;\alpha, \beta)]^2\,,$$ which for the chosen angles gives
$F_\epsilon(\frac{\pi}{4},\frac{\pi}{4})$ equal to QFI. Then we have
to express these optimal estimators,
$\hat\epsilon=V(\pm\pi/4,\pm\pi/4)$, in terms of the coincidences
counts, which are the results  of the measurement process. This can
be done by fixing for example $\alpha=\beta=-\pi/4$
and then, for each measurement run $j=1,..,M=40$, one
records the vector
$\mathbf{k}_{j}=\{k_{0,j},k_{1,j},k_{2,j},k_{3,j}\}$, where
$k_{x,j}\equiv k_{x,j}(-\pi/4,-\pi/4)$, is the number of coincidence
counts for the projector $\Pi_x$ defined in Eq. (\ref{Eq.: projectors
operators})
as measured by the coincidence circuit during a single
time window  of $10$ seconds, and whose expected distribution
is given $$p_x(\epsilon; \alpha,\beta)=\hbox{Tr}[\varrho\,
\Pi_x(\alpha,\beta)]\,.$$
Finally, we have to derive the
probabilities $p_x(\epsilon;-\pi/4,-\pi/4)$ in the
expression of $V(\alpha,\beta)$ in terms of the relative frequencies
$k_{x,j}(\alpha,\beta)/K_j$, where $K_j=\sum_x k_{x,j}$ is the total
number of  coincidences. For large values of $K_j$ the coincidence
rates $k_{x,j}(\alpha,\beta)/K_j$ converges to the probability
$p_x(\epsilon;\alpha,\beta)$. Therefore, the optimal estimator can
be written as desired in terms of the coincidences' vector:
$\hat\epsilon \equiv \hat\epsilon(\bf{k}_j)$.
\par
A second statistical model, which is a possible candidate to represent the output
of our experiment, is the Werner model of Eq. (\ref{Eq.: DM.werner}).  From the
physical point of view it corresponds to incorporate in our our scheme a
portion of ``fake'' coincidences that results from dark counts of SPADs and
from the influence of the ambient unpolarized luminescence. Since this light is
unpolarized, its density operator can be described by the identity in
(\ref{Eq.: DM.werner}). The distribution of coincidences is given by
$$p_x^\prime(\epsilon; \alpha,\beta)=\hbox{Tr}[\varrho^\prime\,
\Pi_x(\alpha,\beta)]\,,$$ and the unbiased estimators for the mixing
parameter and the entanglement negativity of the state by
\begin{align}
\hat p^\prime &= V(0,0) \notag \\
\hat \epsilon^\prime &= -\frac12 + \frac12 V(0,0) + V(-\pi/4,-\pi/4)
\,.\label{estimatorsWerner}
\end{align}
where $V(0,0)=V(\alpha=0,\beta=0)$ has been defined above. The estimators may
be then written in terms of the coincidence vectors $\mathbf{k}_j$, which was
previously defined and that is used for $V(-\pi/4,\pi/4)$, and
$\mathbf{r}_j=\{r_{0,j},r_{1,j},r_{2,j},r_{3,j}\}$, which is used in an
analogous way to define the probabilities in for $V(0,0)$ and whose elements
are defined as $r_{x,j}\equiv r_{x,j}(0,0)$ i.e., the number of coincidence
counts for the projector $\Pi_x$  (\ref{Eq.: projectors operators}) with
$\alpha=0,\beta=0$; in this case the total number of coincidences is
$R_j=\sum_x r_{x,j}$. The estimators can then be written as $\hat p^\prime=\hat
p^\prime(\mathbf{r}_j)$, and $\hat \epsilon^\prime =\hat
\epsilon^\prime(\mathbf{k}_j,\mathbf{r}_j)$.
\section{Results} \label{sec:results}
We first observe that for $\hat\epsilon(\bf{k}_j)$ and finite $K_j$s the
uncertainty in the estimation of the entanglement are mostly due to
fluctuations $\delta k_x$ in the coincidence counts $k_{x,j}$ around
their average values $\average{k_x}=\sum_j k_{x,j}/M$. Thus, if we want
to establish under which conditions on the fluctuations $\delta k_x$ the
variance of the estimator $\hat\epsilon(\bf{k}_j)$ satisfies the
required bound, we have to implement standard uncertainty propagation
with the derivatives $\partial_x \equiv \partial/\partial k_x$ evaluated
for $k_x\equiv \average{k_x}$, and assuming independence among
fluctuations at different angles, we have
\begin{eqnarray}
\hbox{Var} (\hat\epsilon)&=& \sum_x |\partial_x\hat\epsilon|^2 \delta k_x^2
\notag \\ &=& \frac{4}{\average{K}^4}\Big[\big(\langle k_0\rangle+\langle k_3
\rangle\big)^2 \big(\delta k_1^2+\delta k_2^2\big)  \notag \\
&& +\big(\langle k_1\rangle\label{vx}
+\langle k_2\rangle\big)^2\big(\delta k_0^2+\delta k_3^2\big)\Big]
\,. \end{eqnarray}
If we now assume
that the counting processes have a Poissonian statistics, i.e. $\delta
k_x^2=\hbox{Var}(k_x)= \average{k_x}^2$, then it is straightforward
to prove that
$$
\hbox{Var}(\hat\epsilon) = \frac{4}{\average{K}^3}\,
(k_0+k_3)(k_1+k_2)= \frac{1}{\average{K}}\,(1-\hat\epsilon^2)$$ i.e. QC
measurements allow for optimal estimation of entanglement with precision
at the quantum limit.
Since the inverse of QFI is given by
$[H^{-1}]_{\epsilon\epsilon}=1-\epsilon^2$ for a wide range of two-qubit families
of states \cite{EE08}, the above calculations suggest that this is a
general result. In particular, following the discussion at the end of
section \ref{sec:models.theory}, the above result is true also for the
Werner state. In other words, given a source emitting polarization
two-qubit states with coincidence counting statistics satisfying the
Poissonian hypothesis, then the experimental setup of Fig.~\ref{f:setup}
allows for optimal estimation of entanglement at the quantum limit by
means of a QC estimator.
We finally note that in order to test the Poissonian hypothesis in our
experiment we evaluated the Fano factor, which is defined as
$F=\frac{\sigma_\tau^2}{\mu_\tau},$ where $\sigma_\tau^2$ is the
variance and $\mu_\tau$ is the mean of a random process in some time
window $\tau$. For a Poissonian process Fano factor should be equal to
unity. In our experiment we had slightly different values~\cite{EEE},
but the method still allows for optimal estimation,
thus showing the robustness of optimal measurement against deviation from
Poissonian behaviour.
\subsection{Almost pure states}
The experimental setup of Fig.~\ref{f:setup} allows for the preparation of
quantum states with high value of purity, namely having mixing
parameter $p$ close to unity. In these conditions both family of states
in Eq.s (\ref{Eq.: DM.deco}) and (\ref{Eq.: DM.werner})
described in section \ref{sec:models.theory} are expected to give a
reliable estimation of entanglement. In order to verify this assessment,
in the first part of our experiment we have performed measurements with
different values of initial entanglement corresponding to different
values of $q$, i.e. of the angle $\phi$ determined by WP0.  We first consider the
decoherence model of Eq. (\ref{Eq.: DM.deco}).  This model can be
considered as a description of the decoherence mechanisms occurring in
the experimental setup due to fluctuations of the relative phase between
the two polarization components, which results in fluctuation of phase
shift between biphoton created in two crystals. Our experimental procedure is based on
$M=40$ repeated acquisitions of coincidence vector $\boldsymbol{k}_j =
\{k_{0j},k_{1j},k_{2j},k_{3j}\}$.  We have randomized the composition of
$\boldsymbol{k}_j$ over the sequence of measurements to avoid spurious
correlations, and finally we have estimated entanglement as the sample
mean $\langle\hat\epsilon\rangle = \sum_j
\hat\epsilon(\boldsymbol{k}_j)/M$. The corresponding uncertainty has
been evaluated by the sample variance $\hbox{Var}(\hat\epsilon) = \sum_j
[\hat\epsilon(\boldsymbol{k}_j)-\langle\hat\epsilon\rangle]^2/(M-1)$.
\par
\begin{figure}[h]
\includegraphics[width=0.49\columnwidth]{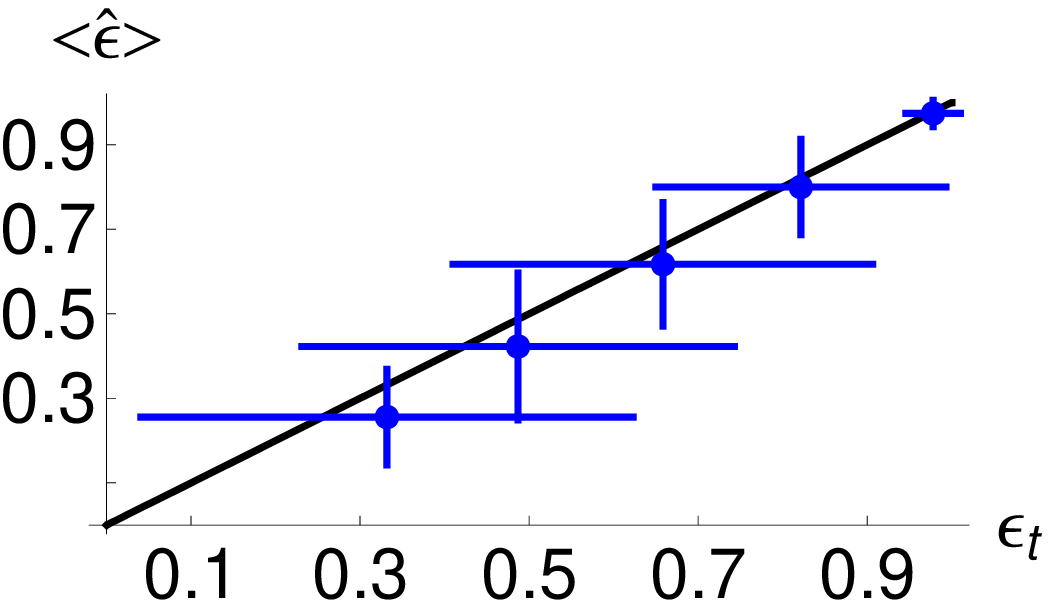}
\includegraphics[width=0.49\columnwidth]{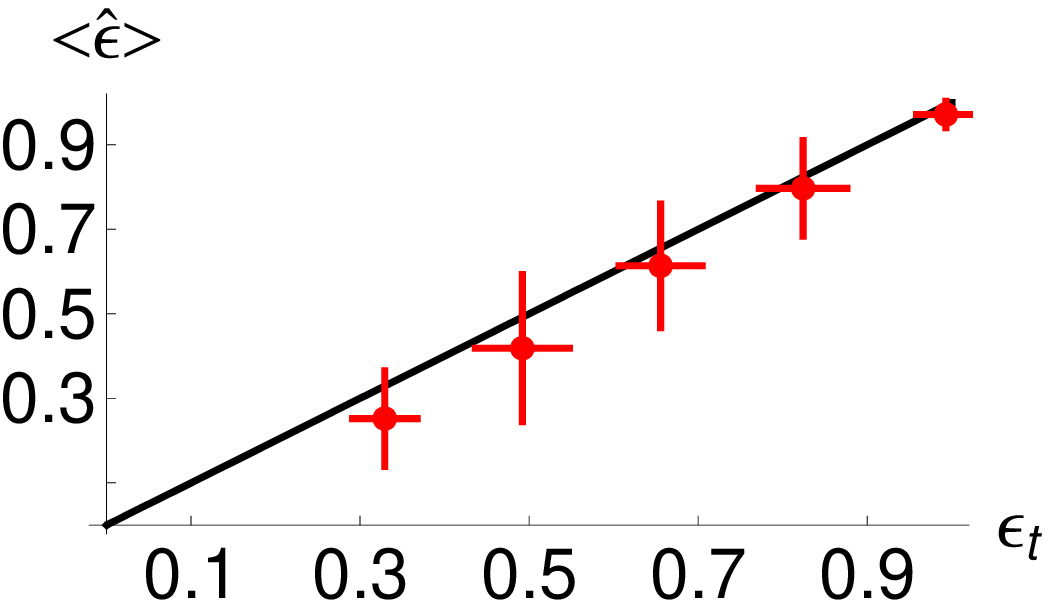}
\caption{(Color online) Estimated value of entanglement as a function of the
reference one assuming, for the description of the output signals, the families
$\varrho$ (left plot) and $\varrho^\prime$ (right plot). The
uncertainty bars stays for the $3\sigma$ confidence interval.}
\label{f:comparison}
\end{figure}

In order to verify the compatibility of data with the decoherence model of
Eq. (\ref{Eq.: DM.deco}) we need to estimate the negativity with a second
procedure, namely we make use of the estimation of the parameter $p$,
quantifying the amount of mixing introduced by decoherence processes. We
therefore define an unbiased estimator $\hat p$ by first reversing
formula of the negativity i.e., $p=\frac12\epsilon/\sqrt{q(1-q)}$.  We
then note that the values of $q$ and $1-q$ in this model are given by
the probabilities relative to the projective measurements $\Pi_0(0,0)$
and $\Pi_3(0,0)$ respectively, that can be expressed in terms of the
elements $r_{0,j}$ and $r_{3,j}$.  The estimator for $p$ then reads
$$\hat p (\boldsymbol{r}_j,\boldsymbol{k}_j)=
\frac12\hat\epsilon(\boldsymbol{k}_j)
\frac{R_j}{\sqrt{r_{0,j}r_{3,j}}}\,,$$ where again $R_j=\sum_x r_{x,j}$.
Rewriting the negativity defined in Eq. (\ref{Eq.: NegativityDefinition})
in terms of the pump
polarization angle $\phi$ we obtain $\epsilon = p \sin 2\phi$.
Thus the reference value $\epsilon_t$ of the negativity is then inferred as
$\epsilon_t=\langle \hat p\rangle\:\sin{2\phi}$, i.e. using the
knowledge of $\phi$ and the
estimation $\average{p}$ of the mixing parameter.
By making use of the relations in Eq. (\ref{estimatorsWerner}) one can
apply the same arguments to the Werner case and derive an appropriate
expression for $\epsilon_t$. Upon evaluating the corresponding sample
means and variances we can therefore obtain the first result of our
analysis. This is illustrated on Fig.~\ref{f:comparison} where we report
the estimated value of entanglement as a function of the reference one
assuming, for the description of the output signals, the families
$\varrho$ (left plot) and $\varrho^\prime$ (right
plot) respectively. Here the uncertainty bars denote the $3\sigma$ confidence
interval and from this plots it is apparent that the experimental data
are compatible with both models.
\par
Notice that the reference value is built, on the basis of a given model,
in part with informations coming from the experimental settings (the
tuning of the angle $\phi$) and in part from the results of suitably
chosen coincidence measurements.  On the other hand, the estimated value
of entanglement is obtained solely with experimental quantities. In
principle, we are not expecting the reference value to be more precise
that the estimated one.  The idea here is to use two different estimates
of the same quantity (entanglement) obtained in two different and
independent ways in order to to discriminate and validate the different
statistical models. Following our analysis, a given model is not
suitable for the description of our system if the two different
estimates that can be derived by that model, together with the resulting
errors, are not compatible.  
\par
It is interesting to compare these results, in particular the ones which refer
to the decoherence model (left plot in Fig.~\ref{f:comparison}), with those
obtained for a different set of measurements data presented in \cite{EEE}. In
that case a less precise control of the temperature of the PDC generation
system made more relevant the fluctuation of the phase and thus the state more
mixed. Therefore, in that case, a self-consistent statistical analysis of the
acquired data allowed discriminating between the two statistical models
identifying the decoherence model of Eq. (\ref{Eq.: DM.deco}) as the correct
one for the experimental set up used in \cite{EEE}. In the present case, which
includes that the already mentioned control in temperature, the states obtained
are nearly pure and thus one cannot expect the different characterization of
noise to be relevant. Furthermore, to experimentally obtain more pure state one
should reduce the collection angle of PDC emission. This obviously reduces the
rate of coincidence counts, thus inducing an increase of the variance of both the
estimators, for negativity and purity parameter respectively.
\begin{figure}[h]
\includegraphics[width=0.49\columnwidth]{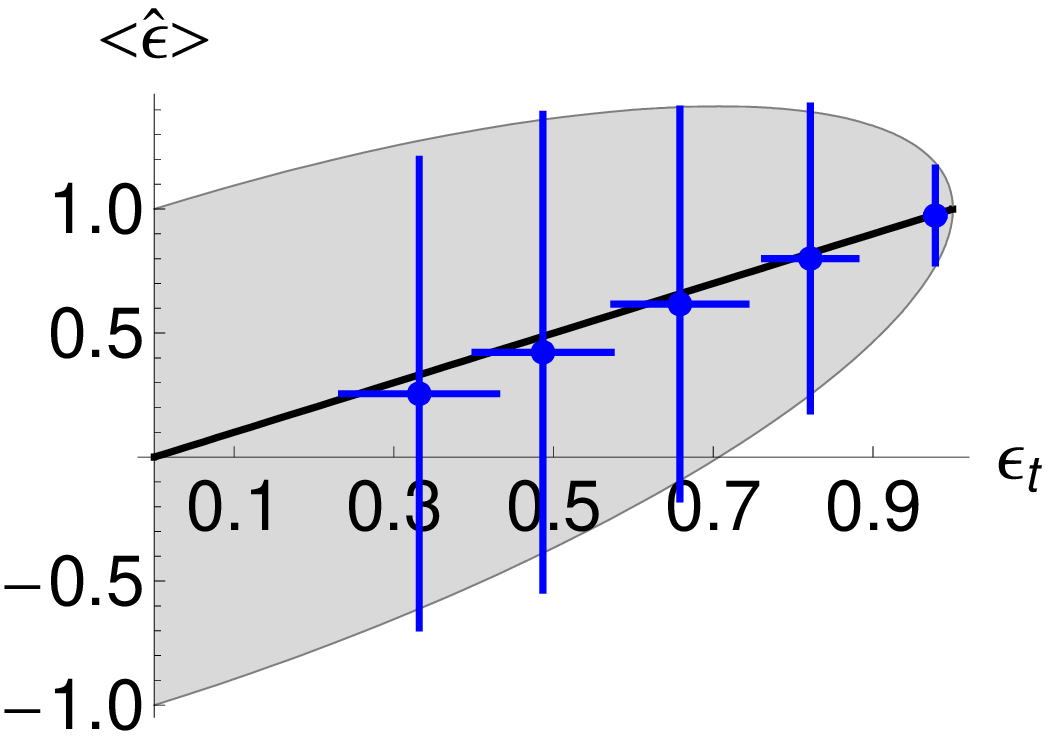}
\includegraphics[width=0.49\columnwidth]{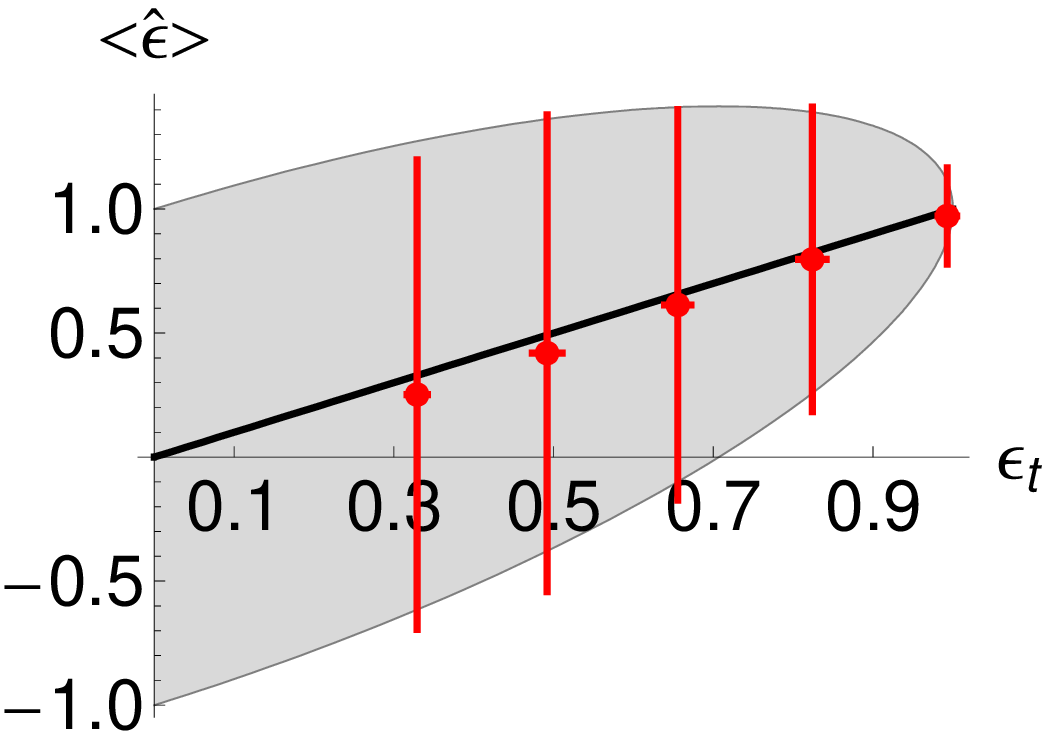}
\caption{(Color Online) Estimation of entanglement at the quantum limit. The
plot shows the estimated value of entanglement $\langle\hat\epsilon\rangle$
according to the decoherence (left) and Werner (right) models as a function
of the reference one $\epsilon_t$. The uncertainty bars on
$\langle\hat\epsilon\rangle$ denotes the quantity
$\sqrt{\hbox{Var}(\hat\epsilon) \times \langle K\rangle}$, i.e. the square root
of the sample variance multiplied by the average number of total coincidences
$\langle K \rangle$. The gray area corresponds to values within the inverse of
the quantum Fisher information $\epsilon_t \pm H_{\epsilon_t}^{-1/2}$. Uncertainty
bars on the abscissae are due to fluctuations in the estimation of the mixing
parameter.}
\label{f:VarQFI}
\end{figure}
\par
We now pass to evaluate the optimality of our estimation procedure.  In
Fig.~\ref{f:VarQFI} we show, for the decoherence (left) and Werner (right)
model, the estimated value of entanglement as a function of the reference one
obtained for different values $q=0.97, 0.93, 0.88, 0.78, 0.5$ (i.e.
$\phi=10^\circ, 15^\circ, 20^\circ, 28^\circ, 45^\circ$). Note that the
corresponding estimated mixing parameter in both model is larger than $0.97$ for all
points. The uncertainty bars on $\langle\hat\epsilon\rangle$ denotes the
quantity $\sqrt{\hbox{Var}(\hat\epsilon) \times \langle K\rangle}$, i.e. the
square root of the sample variance multiplied by the average number of total
coincidences $\langle K \rangle$. This is in order to allow a direct comparison
with the Cramer-Rao bound in term of the inverse of the Fisher information (the
gray area). Uncertainty bars on the abscissae correspond to fluctuations
$\delta\epsilon_t$ in the determination of $\epsilon_t$, due to fluctuations in
the estimation of the mixing parameter with the procedure outlined above. The
plot shows that our procedure allows estimating the entanglement with a
precision at the quantum limit for any value of $q$. From the figure it is also
apparent that, due to the high purity achieved with the experimental set up
that includes the active temperature control, and, therefore, due to the
irrelevance of the decoherence introduced, both the models give optimal
estimation. Notice that this conclusion is robust against the fact that the
statistics is not exactly Poissonian.
\subsection{Comparison with tomographic estimation}
\begin{figure}[h]
\includegraphics[width=0.9\columnwidth]{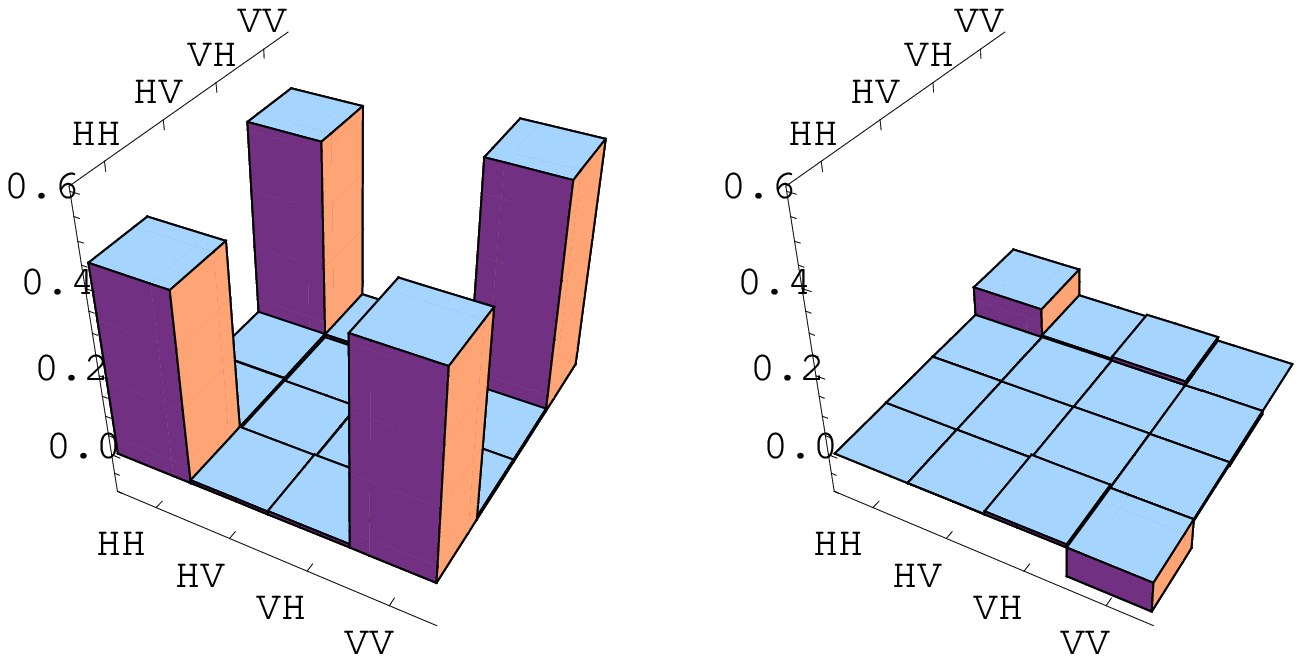}
\includegraphics[width=0.9\columnwidth]{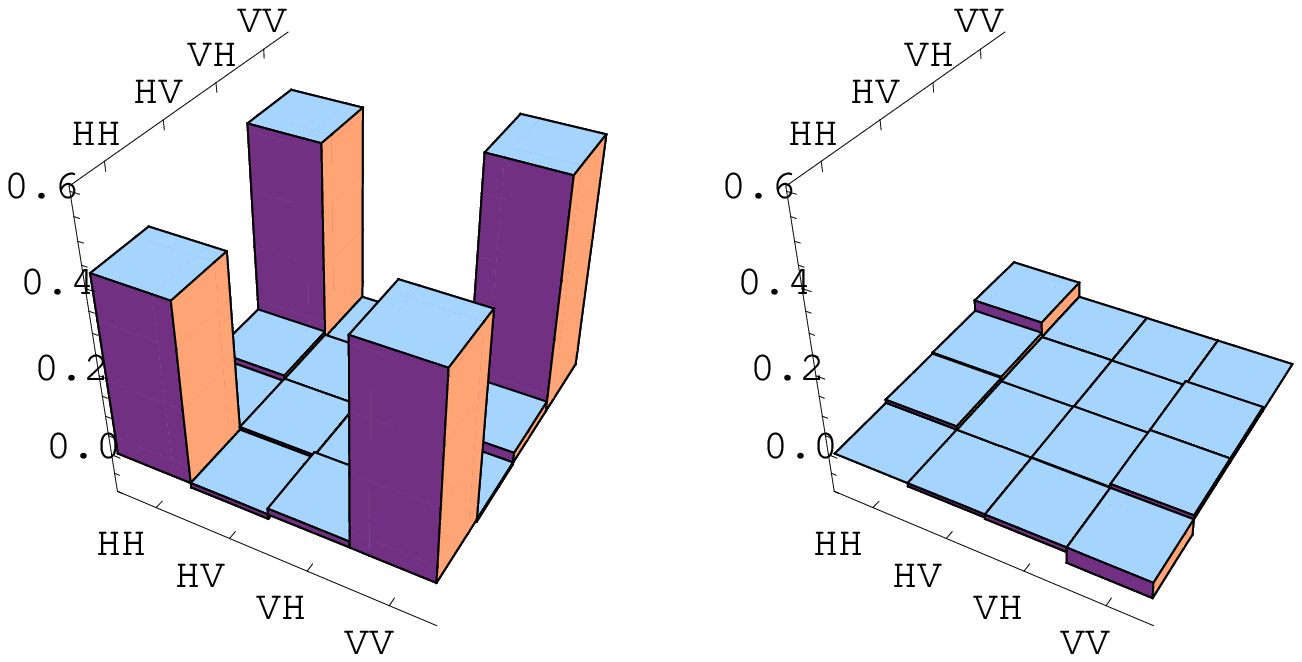}
\caption{(Color online) Real (left) and Imaginary (right) part of
the tomographically reconstructed density matrix for the maximally
entangled state with J16 (top) and R16 (bottom) protocols. All the real
elements, except the four dominant, and the imaginary ones are compatible
with zero within the estimated tomographic uncertainties (not shown
in the figure). }
\label{f:tomo}
\end{figure}

We compared our results with estimation of entanglement from density matrix
elements obtained exploiting two different procedures of quantum state
tomography \cite{KB00,Ja01}. We found that the reconstructed density matrices
are, for both tomography protocols, statistically compatible within with
both the two models of Eqs. (\ref{Eq.: DM.deco}) and (\ref{Eq.: DM.werner}).
As an example we present in Fig.~\ref{f:tomo} real and imaginary part of
reconstructed density matrices of maximally entangled state corresponding to
$q=\frac12$ (i.e., $\phi=45^\circ$).
\par
In fact, the tomographic procedure also allowed us to estimate entanglement
and the corresponding variance. In order to have a fair comparison of the
uncertainties obtained with different methods we have set measurement
time for the tomographic reconstruction such to have thr total number
of registered coincidences counts equal to $M\langle K \rangle$, i.e.
the total number of coincidence in the optimal measurement.  The values
of negativity calculated directly using the reconstructed density matrices
and its variance (obtained by error propagation) for the maximally
entangled state are presented in Table \ref{t:neg} together with the
determination obtained from the optimal measurement maximizing the QFI.
All three negativity values overlap in their uncertainty intervals;
the three methods are therefore coherent. Furthermore, it is evident
from the presented results that the optimal method devised in this paper
allows, at fixed sample size, for a sensitive reduction of the
uncertainty in entanglement estimation.
\begin{table} [h!]
\begin{tabular}{lccc}
 \hline%
 Method & $\epsilon$ & & $\delta\epsilon$ \\
\hline%
Optimal & 0.972 & $\pm$ & 0.011 \\
Tomography: J16 \hspace{0.7cm} & 0.984 & $\pm$ & 0.048 \\
Tomography: R16 \hspace{0.7cm} & 0.957 & $\pm$ & 0.046 \\
\hline%
\end{tabular}
\caption{Estimated value of entanglement with different methods. The
uncertainty $\delta\epsilon$ is calculated usinv Eq. (\ref{vx})
for the optimal method and with error propagation for
tomographic estimation.}
\label{t:neg}
\end{table}
\subsection{Statistical mixtures}
In order to check our method in different working regimes we applied the
estimation procedure to a set of mixed states obtained in a controlled
way, i.e. by adding some portion of unentangled light to pure
entangled state. As we have described in the previous section, our
experimental set up allows us to obtain states with an extremely high
purity. In the following we thus assume that the output state of our
apparatus is the pure state as in Eq. (\ref{Eq.: state_with_q}). Then, if
one is able, for example, to mix in a controlled way the components
$\ketbra{HH}{HH}$ and $\ketbra{VV}{VV}$ to the maximally entangled states
one obtains the states
\begin{eqnarray}\label{Eq.: DM.deco.full}
\varrho & =&p\ketbra{\psi}{\psi}+(1-p)\,D \\
D & =& \frac12 \left(\ketbra{HH}{HH} + \ketbra{VV}{VV})\right) \notag\,,
\end{eqnarray}
which correspond to the model (\ref{Eq.: DM.deco}) with an adjustable
mixing parameter. In practice, in order to tune the value of the mixing
parameter $p$ we have measured coincidence counts for states
$\ketbra{\HH}{\HH}$ and $\ketbra{\VV}{\VV}$ for different time intervals.
The sample of coincidence counts is then added to experimental data
obtained for the maximally entangled pure state and then analyzed as in the
previous section.
\begin{figure}[h!]
\includegraphics[width=0.49\columnwidth]{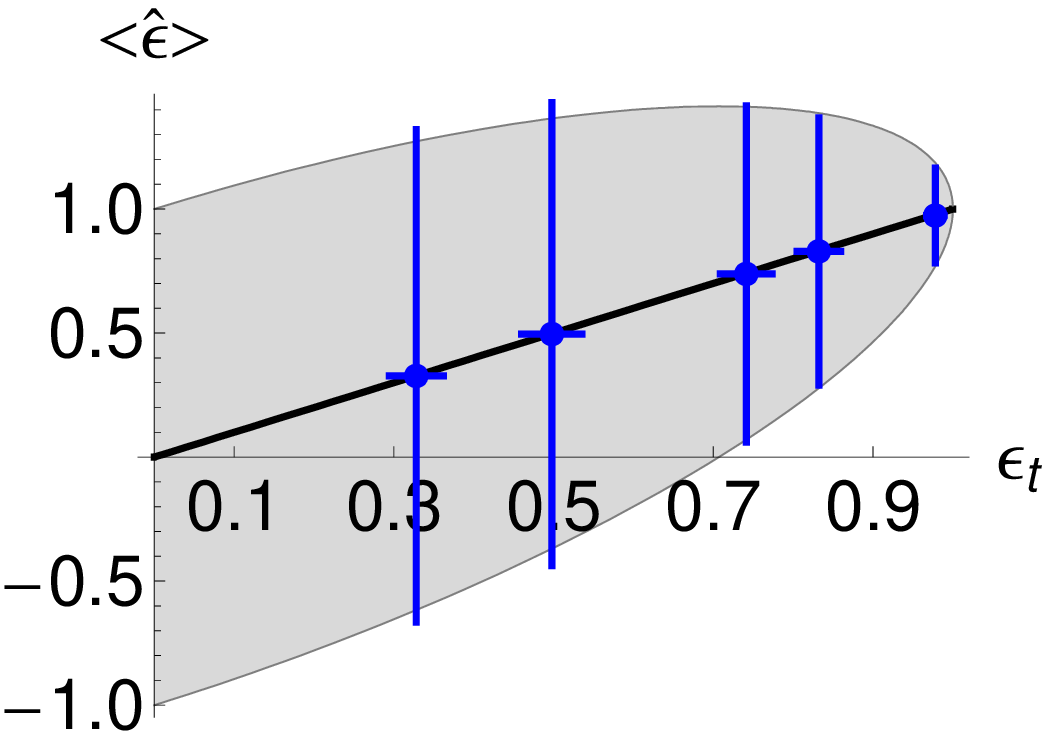}
\includegraphics[width=0.49\columnwidth]{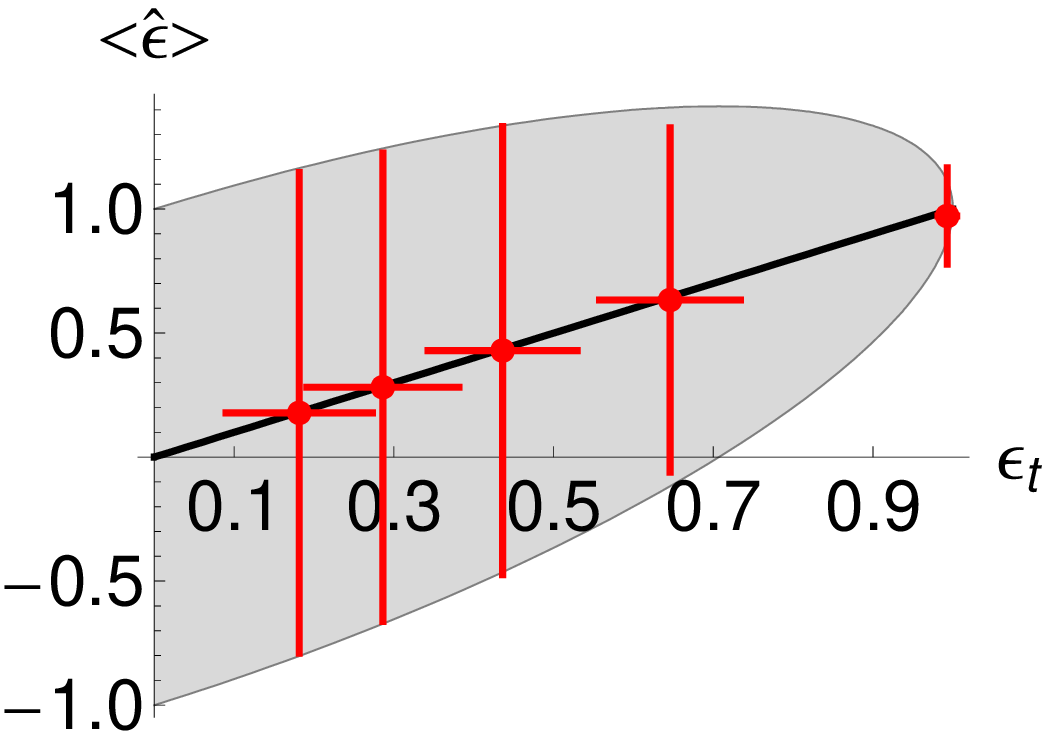}
\caption{(Color Online) Estimation of entanglement at the quantum limit. The
plot shows the estimated value of entanglement $\langle\hat\epsilon\rangle$
as a function of the reference one $\epsilon_t$. In the left panel we report
estimated entanglement for mixed states generated according to the
decoherence model (\ref{Eq.: DM.deco.full}). In the right panel we report
estimated entanglemed for mixed states generated according to the Werner
model (\ref{Eq.: DM.werner.full}). The points correspond to different
portions of incoherent addition from both crystals.} \label{f:MixVarQFI}
\end{figure}
\par
In the left panel of Fig.~\ref{f:MixVarQFI} we show the estimated value
of entanglement as a function of the reference one for the originally maximally
entangled state ($q=\frac12$) and for states prepared with mixing parameter
$p=99.5\%, 83\%, 74\%, 50\%, 33\%$.
\par
A similar analysis may carried out for the Werner model. In this case, in
order to tune the value of the mixing parameter $p$ one
should supplement the coincidences vectors $\bf{k}_j$ and
$\bf{r}_j$ with values coming from unpolarized light.
This can be achieved by measuring coincidence counts for
$\ketbra{\HH}{\HH}$, $\ketbra{\HV}{\HV}$, $\ketbra{\VH}{\VH}$ and
$\ketbra{\VV}{\VV}$ for different time intervals. The measured
values are then added to the previously measured values
for pure maximally entangled state. In this way, one can get
data corresponding to
\begin{eqnarray}\label{Eq.: DM.werner.full}
 \varrho^\prime=p\ketbra{\psi}{\psi}+(1-p) \frac{\mathbbm I}{4}
\end{eqnarray}
which correspond to a Werner state with tunable depolarizing
parameter. After performing measurement and analysis set described
in previous section we can estimate entanglement and mixing
parameter value in this family of states. In the right panel
Fig.~\ref{f:MixVarQFI} we show the estimated value of entanglement
as a function of the actual one for the originally maximally
entangled state and mixture parameter $p=99.5\%,
76\%, 62\%, 52\%, 45\%$. As one can evince from the presented figure
our method provides optimal entanglement estimation also for mixed
states.
\section{Conclusions}
\label{sec:outro} In this paper we have addressed in detail the
estimation of entanglement for pairs of polarization qubits. Our
scheme is based on visibility measurements of quantum correlations
and allows  optimally estimating entanglement of families of
two-photon polarization entangled states without the need of
performing full tomography. Our procedure is self-consistent and
allows estimating the amount of entanglement with the ultimate
precision imposed by quantum mechanics. Although optimal estimation
of entanglement does not require the full tomography of the states
we have also performed state reconstruction using two different sets
of projectors and explicitly shown that they provide a less precise
determination of entanglement.
\par
The technique has been demonstrated for nearly pure states as well as
for controlled mixtures in order to confirm its reliability in any
working regime.  With a suitable choice of correlation measurements it
may be extended to a generic class of two-photon entangled states. The
statistical reliability of our method suggests a wider use in precise
monitoring of external parameters assisted by entanglement.
\section*{Acknowledgements}
This work has been supported by Associazione Sviluppo Piemonte. MGAP
and PG thanks Marco Genoni for several useful discussions. MGAP
thanks Simone Cialdi and Davide Brivio for useful discussions.

\end{document}